\documentclass[twocolumn]{aastex701}
\shorttitle{The BlueDOG at Cosmic Noon: A Possible Analog to Little Red Dots?}
\shortauthors{Kim et al.}
\graphicspath{{./}{figures/}}

\begin{document}

\title{The BlueDOG at Cosmic Noon: A Possible Analog to Little Red Dots?}

\correspondingauthor{Woong-Seob Jeong}
\email{jeongws@kasi.re.kr}

\author[0009-0002-3931-6697]{Seongjae Kim}
\email{seongjkim@kasi.re.kr}
\affiliation{Korea Astronomy and Space Science Institute, 776 Daedeok-daero, Yuseong-gu, Daejeon 34055, Korea}
\affiliation{University of Science and Technology, 217 Gajeong-ro, Yuseong-gu, Daejeon 34113, Korea}

\author[0000-0002-2770-808X]{Woong-Seob Jeong}
\email{jeongws@kasi.re.kr}
\affiliation{Korea Astronomy and Space Science Institute, 776 Daedeok-daero, Yuseong-gu, Daejeon 34055, Korea}
\affiliation{University of Science and Technology, 217 Gajeong-ro, Yuseong-gu, Daejeon 34113, Korea}

\author[0000-0002-3560-0781]{Minjin Kim}
\email{mkim.astro@gmail.com}
\affiliation{Department of Astronomy and Atmospheric Sciences, Kyungpook National University, Daegu 41566, Korea}

\author[0000-0003-1470-5901]{Hyunsung D. Jun}
\email{hyunsungjun@gmail.com}
\affiliation{Department of Physics, Northwestern College, 101 7th St. SW, Orange City, IA 51041, USA}

\author[0000-0003-3078-2763]{Yujin Yang}
\email{yyang@kasi.re.kr}
\affiliation{Korea Astronomy and Space Science Institute, 776 Daedeok-daero, Yuseong-gu, Daejeon 34055, Korea}
\affiliation{University of Science and Technology, 217 Gajeong-ro, Yuseong-gu, Daejeon 34113, Korea}

\author[0000-0002-6660-9375]{Takao Nakagawa}
\email{nakagawa@ir.isas.jaxa.jp}
\affiliation{Institute of Space and Astronautical Science, Japan Aerospace Exploration Agency, 3-1-1 Yoshinodai, Chuo-ku, Sagamihara, Kanagawa 252-5210, Japan}
\affiliation{Advanced Research Laboratories, Tokyo City University, 1-28-1 Tamazutsumi, Setagaya-ku, Tokyo 158-8557, Japan}





\begin{abstract}
We discovered a hyperluminous dust-obscured galaxy with mysterious blue-excess emission (BlueDOG) in rest-frame UV of its spectral energy distribution (SED) from a multi-wavelength survey in the AKARI Deep Field - South (ADF-S). We present the results of SED analysis with multiwavelength photometric data and spectroscopic analysis, observed with Gemini-S/GMOS, FLAMINGOS-2, to explore the origin of blue-excess emission of a hyperluminous BlueDOG, ADFS-KMTDOG-102, at \textit{z}=2.6. The SED analysis shows that this BlueDOG is a highly massive system (log~$M_{*}$/$M_\odot=12.3$) with substantial extinction. Additionally, the proportion of the old stellar population exceeds that of the young stellar population, which suggests stellar evolution cumulated from the early universe. The mass of supermassive black hole (SMBH) estimated using the extinction-corrected broad H$\rm\alpha$ emission line yields log~$M_{\rm BH}$/$M_\odot$=10.2. We discuss the similarity between the BlueDOG and `Little Red Dots' (LRDs), recently discovered with the James Webb Space Telescope, showing SED shapes remarkably similar to those of LRDs. The UV emission line ratios indicate that the emission lines are primarily powered by the central active galactic nuclei (AGN). In contrast, the origin of the blue-excess UV continuum remains ambiguous, since both recent star formation and AGN-induced scattered light are viable explanations, based on the results from the SED fitting and scattered light modeling.

\end{abstract}



\section{Introduction}\label{sec:intro}
It is believed that infrared (IR) luminous galaxies are a crucial population for understanding massive galaxy formation via gas-rich mergers. These IR bright galaxies typically present intense star formation with heavy obscuration by dust and gas, which drives faint UV/optical and bright IR luminosity. These galaxies also exhibit efficient mass accretion onto super massive black holes (SMBHs) (e.g., \citealt{2008ApJS..175..356H}; \citealt{2012ApJ...758L..39T}; \citealt{2018PASJ...70S..37G}). An existence of SMBHs has been confirmed in several dusty populations for example submillimeter galaxies (SMGs; \citealt{2002PhR...369..111B}), dust-obscured galaxies (DOGs; \citealt{2008ApJ...677..943D}), and hot dust-obscured galaxies (hot DOGs; \citealt{2012ApJ...755..173E}) (\citealt{2008AJ....135.1968A} for SMGs; \citealt{2011AJ....141..141M} for DOGs; \citealt{2015ApJ...804...27A}, \citealt{2018ApJ...852...96W} for hot DOGs).

DOGs were originally defined based on their optical and mid-IR (MIR) colors (F$_{24\mu m}$/F$_{R}$ $\gtrsim$ 1000) \citep{2008ApJ...677..943D} to identify dusty and heavily obscured galaxies at high redshift. \citet{2008ApJ...677..943D} showed that DOGs can be classified into two types, bump DOG and power-law DOG. The fraction of power-law DOGs tends to increase with 24 $\mathrm{\mu m}$ flux. Moreover, the 24$\mathrm{\mu m}$-bright power-law DOGs present higher AGN contribution to bolometric luminosity than bump DOGs in a merger simulation \citep{2010MNRAS.407.1701N}.

Recently, new populations of dusty galaxies exhibiting unexpected blue-excess emission at rest-frame UV while showing redder colors at longer wavelengths have been discovered at \textit{z}$\sim$2-3 (BluDOG: \citealt{2019ApJ...876..132N}, Blue Hot DOGs (BHDs): \citealt{2016ApJ...819..111A}; \citealt{2020ApJ...897..112A}). \citet{2020ApJ...897..112A} proposed and investigated three possible scenarios to account for the blue excess emission in BHDs: (1) dual AGN, unobscured faint AGN and obscured AGN, (2) extreme starburst, and (3) leaked UV light from central obscured AGN. They suggested that the blue excess emission of BHDs seems to be scattered light from highly obscured AGN \citep{2022ApJ...934..101A}.

In this study, we present the spectroscopic results for a hyperluminous BlueDOG at $\textit{z}=2.6$. We describe our BlueDOG sample and observations in Section~\ref{sec:data}. In Section~\ref{sec:result}, we show the redshift measurement, line ratios from the observed spectrum, and the SMBH mass measurement. We discuss Ly$\rm \alpha$ escape fraction, implications from emission lines, and Eddington ratio in Section~\ref{sec:discussion}. Throughout this paper, we adopted a concordance cosmology with $H_0=70$\,km\,s$^{-1}$\,Mpc$^{-1}$, $\Omega_\Lambda = 0.7$, $\Omega_M = 0.3$, and AB magnitude system.

\section{Sample and Observations}\label{sec:data}
\subsection{BlueDOG sample}\label{subsec:BlueDOG}
The dust-obscured galaxies (DOGs; \citealt{2008ApJ...677..943D}) in AKARI Deep Field -- South (ADF--S) were selected using multi-wavelength data ranging from optical to sub-mm wavelengths, including newly acquired optical images from Korea Microlensing Telescope Network (KMTNet) \citep{2016JKAS...49..225J}. We targeted hyperluminous DOGs with $L_{\rm bol}>10^{13}L_{\odot}$ through near-infrared (NIR) spectroscopic observations. Among our spectroscopic follow-up samples, we identified one hyperluminous DOG (target name: ADFS-KMTDOG-102) exhibiting significant blue-excess emission in rest-frame UV of its spectral energy distribution (SED). Here, we refer to this peculiar DOG population as BlueDOG. The hot DOGs and DOGs having blue-excess emission were reported by \citet{2016ApJ...819..111A, 2020ApJ...897..112A} and \citet[][called as BluDOG]{2019ApJ...876..132N, 2022ApJ...941..195N}, respectively. This BlueDOG also meets the criterion of blue-excess DOGs (i.e., $f_{\nu}\propto \lambda^{\alpha_{\mathrm{opt}}}$ where $\alpha_{\mathrm{opt}}<0.4$ for an optical continuum slope in the observed frame) as defined by \citet{2019ApJ...876..132N}. The power-law index of our target is estimated to be $\alpha_{\rm opt}$=0.036 from optical g-band to y-band of the Dark Energy Survey second Data Release (DES DR1; \citealt{2018ApJS..239...18A}). 

\subsection{Spectroscopic Observations and Data Reductions}\label{subsec:observation}
To investigate the origin of blue excess emission, we performed the spectroscopic follow-up observations for one of the BlueDOG candidates, ADFS-KMTDOG-102, using FLAMINGOS-2 and Gemini Multi-Object Spectrographs (GMOS) on the Gemini-South 8.1m telescope at Cerro Pachon, Chile. Here, we describe the spectroscopy configuration and data reduction for each observation.

\subsubsection{Gemini-South/FLAMINGOS-2 long-slit}\label{subsubsec:gemini_f2}
The NIR long-slit spectroscopic data for ADFS-KMTDOG-102 was obtained using FLAMINGOS-2 on November 22, 2017. HK grism with a slit width of 0.54 arcsec (3 pixels) was used to achieve a spectral resolution of R$\sim$800 with wavelength coverage from 1.2 to 2.4 $\mu m$. The total exposure time was 1200 seconds, with a single exposure time of 300 seconds. We adopted ABBA nodding to subtract the sky background properly. Also, we observed an A0V star (HIP 18384) as a telluric standard with ABBA nodding and 5 seconds of a single exposure. We reduced the raw files using \texttt{Gemini IRAF}, performing bias correction, dark subtraction, flat fielding, wavelength calibration, 2D spectrum combine, and 1D spectral extraction. The 2D spectrum was combined as an average to improve S/N. The \texttt{Xtellcor\_general} \citep{2003PASP..115..389V} was used for telluric correction and flux calibration. 

\subsubsection{Gemini-South/GMOS long-slit}\label{subsubsec:gemini_gmos}
Additionally, optical long-slit spectroscopy was performed with GMOS on December 4, 2020. We adopted a B600/G5323 grating with a slit width of 0.75 arcsec. No order-blocking filter was used in order to cover widely the rest-frame UV emission, which is crucial to unveil the nature of the target. $2\times 2$ binning along both spatial and wavelength directions was applied to enhance the S/N of the spectrum. 
We set a central wavelength of the grating to 470\,nm and 480\,nm to implement wavelength dithers to compensate for spectrum loss by GMOS chip gaps. The total exposure time is 8,400 seconds with 28 frames corresponding to a single exposure time of 300 seconds. Half of the frames have a central wavelength of 470\,nm and the other half has a central wavelength of 480\,nm. The observations were conducted in staring mode rather than nodding mode. For flux calibration, the standard star of the \textit{Hiltner 600}  was observed with the same configuration. Similar to the data reduction of FLAMINGOS-2, \texttt{Gemini IRAF} was used for bias correction, dark subtraction, flat fielding, wavelength calibration, and 2D spectrum combination. 

We used different apertures to extract the 1D spectra for each emission line, ensuring accurate flux preservation. Only Ly$\mathrm{\alpha}$ emission originates from a spatially extended region with a size of approximately 2 arcsec. Then, we combined the Ly$\mathrm{\alpha}$-side spectra with 1D spectra from other spectral regions.

\begin{figure}
    \includegraphics[scale=0.63, trim={1.5cm 0 0 0cm}]{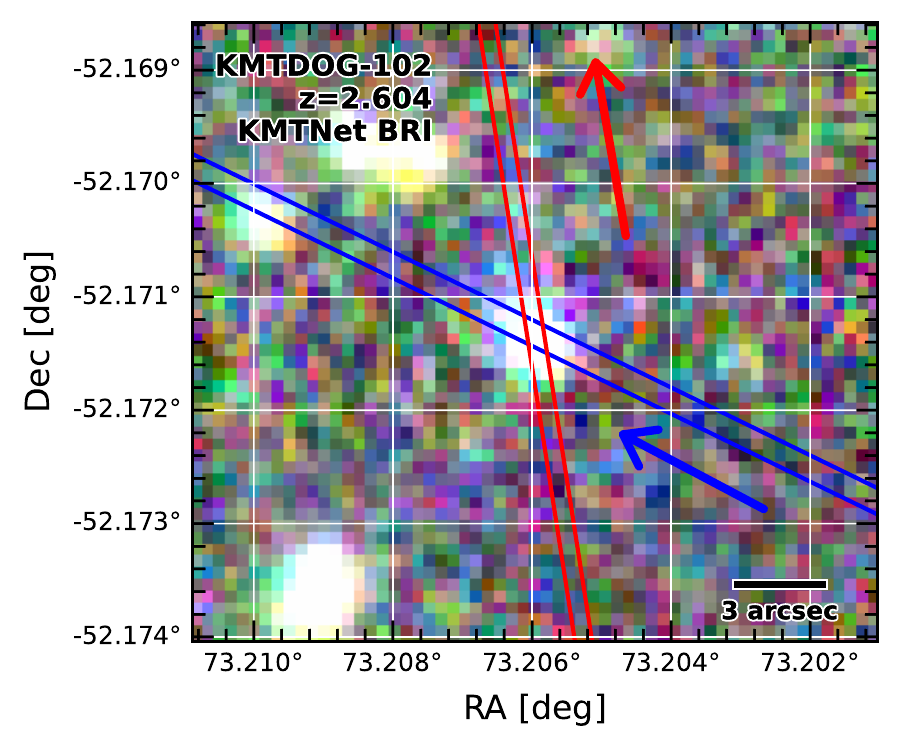}
    \caption{Optical color composite image using KMTNet $B$, $R$, and $I$ bands. The positions of longslits from GMOS (blue) and FLAMINGOS-2 (red) are overlaid with the slit width corresponding to each instrument. In the GMOS observation, we aligned the slit to include the center of the BlueDOG and the blue-excess region where the irregular structure is prominent in the B-band image. Blue and red arrows represent the positive spatial direction of the 2D spectrum for GMOS and FLAMINGOS-2, respectively. \label{fig:KMTNet_image_slit}}
\end{figure}

\begin{figure*}[t!]
    \includegraphics[scale=0.42, trim={2.0cm 0.5cm 0 0cm}]{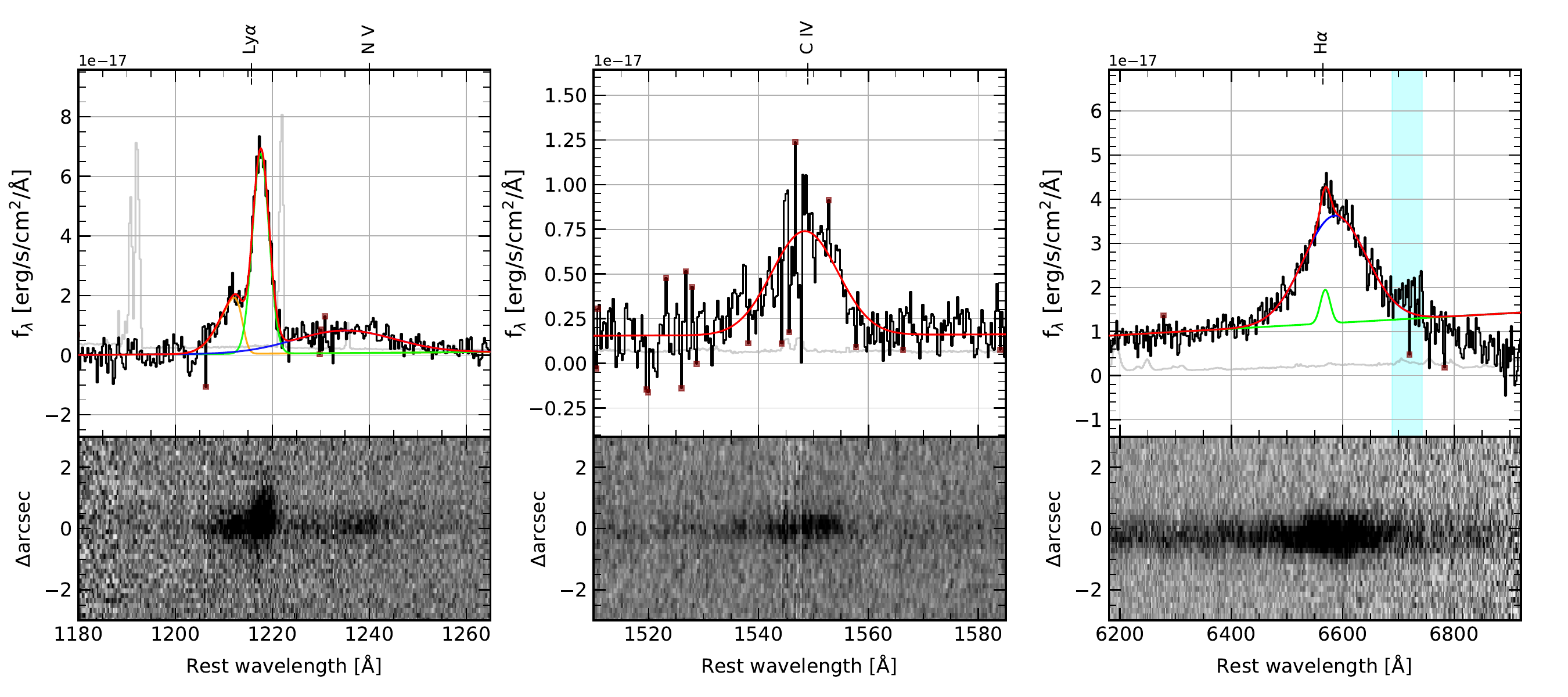}
    \caption{Rest-frame 1-D and 2-D spectrum for the BlueDOG, ADFS-KMTDOG-102, from Gemini-S/GMOS (left and center) and Gemini-S/FLAMINGOS-2 (right). The spectrum and error are plotted as black and gray color in the three panels, respectively. The locations of detected lines are marked with dashed lines and texts. The spiky points with a greater difference than 3$\sigma$ are clipped, which are shown in brown. (left) The best fitting results of double-peaked Ly$\rm{\alpha}$ and broad \ion{N}{5} are presented in orange, green solid lines for Ly$\rm{\alpha}$ and blue solid line for \ion{N}{5}. The red solid line is the summed result of all fitted components. Note that only Ly$\rm{\alpha}$ shows spatially extended shape compared to all observed emission lines. (center) The \ion{C}{4} emission at the line center is affected by a sky emission, [\ion{O}{1}] $\lambda\lambda 5577$ airglow. (right) Broad and narrow components of H$\rm{\alpha}$ are plotted in blue and green solid lines, respectively. In the continuum determination, we omit the continuum at $\lambda_{\rm{rest}} \gtrsim 6800\rm{\AA}$ since that region is located at the edge of FLAMINGOS-2, corresponding observed wavelength of 2.45$\,\rm{\mu m}$. Note that the far-end wavelength coverage of FLAMINGOS-2 is $\sim$2.5$\,\rm{\mu m}$. The cyan shaded region shows a mask for [\ion{S}{2}]$_{6716,\,6731}$ to prevent overestimating the black hole mass.\label{fig:spectrum}}
\end{figure*}

\section{Results}\label{sec:result}

\subsection{Redshift Measurement}\label{subsec:redshift_measurement}
To estimate the redshift of ADFS-KMTDOG-102, we first concatenate the optical and the NIR spectra, masking sky emission lines and telluric regions. Because strong AGN lines (e.g., [\ion{O}{3}] $\lambda\lambda 4959,\ 5007$), commonly used to estimate the redshift, were not properly obtained due to the telluric absorption, the cross-correlation method between the composite spectrum of SDSS quasars spanning from rest-frame UV to NIR (\citealt{2001AJ....122..549V}; \citealt{2006ApJ...640..579G}) and our spectrum was applied, as demonstrated by  \citet{2017NatSR...741617K}. Before this process, we subtracted the continuum and normalized the spectra at the peak of Ly$\mathrm \alpha$ for both spectra, because the slope of the continuum in ADFS-KMTDOG-102 is slightly milder than those in the SDSS composite spectrum. The determined redshift of ADFS-KMTDOG-102 is $2.604\pm 0.026$. 

\subsection{Analysis of NIR and optical emission lines}\label{subsec:emission_line_fitting}
The reduced 1D and 2D spectra of ADFS-KMTDOG-102 in the rest frame are presented in Figure~\ref{fig:spectrum} (Jeong et al. in prep for full DOG samples). We fitted H$\mathrm \alpha$ emission using two (broad and narrow) Gaussian components after masking two [\ion{S}{2}] lines, constraining the FWHM of the narrow component to be less than $1200$ km s$^{-1}$ to reflect the host galaxy's kinematics.  
The best-fit parameters for the H$\mathrm\alpha$ emission line are listed in Table~\ref{tab:line_fitting_param}. As the spectral regions around the H$\beta$ and [\ion{O}{3}] lines are severely degraded by telluric absorptions, reliable spectral measurements for those lines are not available.

As shown in Figure~\ref{fig:spectrum}, Ly$\mathrm \alpha$, \ion{N}{5}, and \ion{C}{4} emission lines are identified in the GMOS spectrum. Interestingly, unlike other rest-frame UV emission lines, the Ly$\mathrm \alpha$ emission extends spatially up to $\sim$2 arcsec in the 2D spectra. We fitted Ly$\mathrm \alpha$ emissions with two Gaussian components to account for the blue-shifted feature, while \ion{N}{5} is fitted with a single Gaussian component. The velocity separation between the two Ly$\mathrm \alpha$ components is approximately $1,360$~km/s, which is significantly larger than previously observed values of 220 $\pm$ 20~km/s in Lyman-alpha Emitters (LAEs) (see Figure~10 in \citealt{2018A&A...619A.136M}). \ion{C}{4} emission line is a broad single Gaussian. The spectral measurements for the emission lines detected at the rest-frame UV are also summarized in Table~\ref{tab:line_fitting_param}.  Note that FWHMs are corrected for the instrumental resolution. The observed flux ratios of H$\mathrm \alpha$/Ly$\mathrm \alpha$, Ly$\mathrm \alpha$/\ion{N}{5}, and \ion{N}{5}/\ion{C}{4} are 9.03$\pm0.35$, 2.04$\pm 0.15$, and 2.26$\pm 0.18$, respectively.

\startlongtable
\begin{deluxetable*}{lccccc} 
    \colnumbers
    \tabletypesize{\scriptsize}
    \tablewidth{0pt} 
    \tablenum{1}
    \tablecaption{Measurements of Emission Lines from optical and NIR spectrum \label{tab:line_fitting_param}}
    \tablehead{
    \colhead{Line} & \colhead{Component} & \colhead{FWHM$_{\rm rest}$ [$\mathrm{km~s^{-1}}$]}& \colhead{Flux [$\mathrm{erg~s^{-1}~cm^{-2}}$]} & \colhead{Luminosity [$\mathrm{erg~s^{-1}}$]} & \colhead{EW$_{\rm rest}$ [{\AA}]}
    } 
    \startdata
    {}& {Blue peak} & 1348.6$\pm 153.8$ & (3.99$\pm 0.28$)$\times 10^{-16}$ & (2.19$\pm 0.15$)$\times 10^{43}$ & 257.9$\pm 18.2$ \\
    {Ly$\mathrm \alpha$}& {Red peak} & 742.8$\pm 47.5$ & (1.04$\pm 0.03$)$\times 10^{-15}$ & (5.73$\pm 0.18$)$\times 10^{43}$ & 564.1$\pm 17.9$ \\    
    {} & {Red$\,+\,$Blue} & \nodata & (1.44$\pm 0.04$)$\times 10^{-15}$ & (7.92$\pm 0.24$)$\times 10^{43}$ & 822.0$\pm 25.5$ \\
    \hline
    {\ion{N}{5}$_{1240}$}& Single & 5835.3$\pm 473.7$ & (7.06$\pm 0.47$)$\times 10^{-16}$ & (3.88$\pm 0.26$)$\times 10^{43}$ & 282.3$\pm 19.3$ \\
    \hline
    {\ion{C}{4}$_{1549}$}& Single & 2685.9$\pm 129.8$ & (3.15$\pm 0.13$)$\times 10^{-16}$ & (1.73$\pm 0.07$)$\times 10^{43}$ & 55.3$\pm 2.3$ \\
    \hline
    { }& Broad & 6055.5$\pm 167.6$ & (1.24$\pm 0.03$)$\times 10^{-14}$ & (6.82$\pm 0.15$)$\times 10^{44}$ & 288.9$\pm 6.3$\\
    {H$\mathrm \alpha$}& Narrow & 890.0$\pm 223.2$ & (6.14$\pm 1.57$)$\times 10^{-16}$ & (3.37$\pm 0.86$)$\times 10^{43}$ & 14.4$\pm 3.7$\\
    { }& Broad$\,+\,$Narrow & \nodata & (1.30$\pm 0.03$)$\times 10^{-14}$ & (7.16$\pm 0.17$)$\times 10^{44}$ & 303.4$\pm 7.3$\\
    \enddata
    \tablecomments{Column (1): Name of emission line, (2): Gaussian component, (3): Rest-frame FWHM in velocity, (4): Line flux, (5): Line luminosity, (6) Equivalent width in rest-frame. Note that only Galactic extinction is corrected for all line fluxes and luminosities, here. However, we use intrinsic extinction corrected H$\alpha$ luminosity for SMBH mass estimation.}
    \end{deluxetable*}

\subsection{Physical Properties from SED Fitting}\label{subsec:sed_fitting}
To derive galaxy properties of the target, the SED fitting was performed with the latest version of Code Investigating GAlaxy Emission (CIGALE; \citealt{2005MNRAS.360.1413B}; \citealt{2009A&A...507.1793N}; \citealt{2019A&A...622A.103B}; \citealt{2022ApJ...927..192Y}).  The updated CIGALE enables us to model stellar, dust, and AGN components, with a recent integration of X-CIGALE (\citealt{2020MNRAS.491..740Y}), additionally allowing the model of X-ray SED and polar dust originated from AGN (\citealt{2022ApJ...927..192Y}). The photometric data points from optical to submm, including  $B$, $R$, $I$ (KMTNet; \citealt{2016JKAS...49..225J}), $g$, $r$, $i$, $z$, $y$ (DES; \citealt{2018ApJS..239...18A}), $H$, $K\rm_{s}$ (VISTA Hemisphere Survey; \citealt{2013Msngr.154...35M}), Ch1 (Spitzer/IRAC; \citealt{2016ApJS..223....1B}), W1, W2, W3, W4 (ALLWISE; \citealt{2014yCat.2328....0C}), MIPS 24$\mu m$, MIPS 70$\mu m$ (Spitzer/MIPS; \citealt{2011MNRAS.411..373C}), FIS 90$\mu m$ (AKARI/FIS; Private comm.), PACS$_{green}$ (Herschel/PACS; \citealt{2012MNRAS.424.1614O}), $PSW$, $PMW$, and $PLW$ (Herschel/SPIRE; \citealt{2014MNRAS.444.2870W}), are used to derive the best-fit SED. The photometric data points are listed in Table~\ref{tab:photometry} for the BlueDOG.

\begin{figure}[h]
    \includegraphics[scale=0.37, trim={2.0cm 0 0 0cm}]{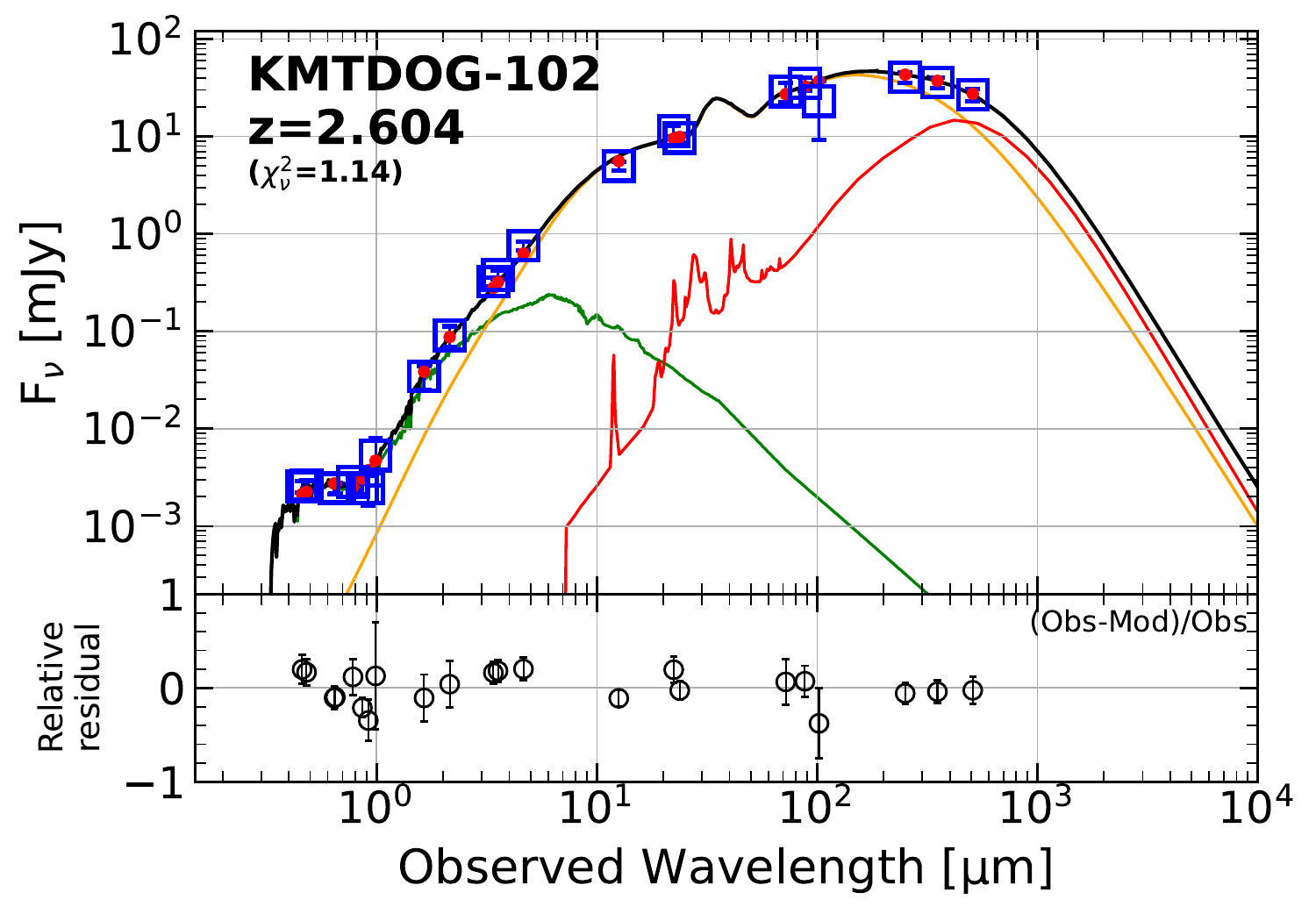}
    \caption{Best-fit SED for the BlueDOG at \textit{z}=2.604, ADFS-KMTDOG-102. The green, orange, and red lines represent attenuated stellar component, AGN component with polar dust, and dust component related to star formation, respectively. The observed fluxes with error bars at each band are plotted in blue squares. Also, the red points are fluxes of the best-fit model at each band. \label{fig:SED}}
\end{figure}

As for the configuration of CIGALE, a delayed star formation history (\textit{sfhdelayed}; \citealt{2015A&A...576A..10C}) with a burst of star-formation, stellar population from \citet{2003MNRAS.344.1000B} with an initial mass function (IMF) of \citet{2003PASP..115..763C}, and nebular emission from \citet{2011MNRAS.415.2920I} are implemented, considering the expected properties of the BlueDOG. The attenuation law modified by \citet{2019A&A...622A.103B} was estimated from starburst galaxies with $R_{V}=4.05$ \citep{2000ApJ...533..682C}, because DOGs are generally thought to be highly dusty and gaseous. $E(B-V)_{stellar}$ corresponding to $0.44E(B-V)_{lines}$ varies from 0 to 1.1 with a step of 0.1. The AGN contribution ($f_{\rm AGN}$) is calculated with the \texttt{SKIRTOR} model \citep{2012MNRAS.420.2756S, 2016MNRAS.458.2288S} by allowing inclination angle, opening angle, and extinction caused by polar dust to vary. Given the presence of broad H$\alpha$ emission, we assumed that the target is viewed face-on with the inclination angle from 0$-$30\,deg. $E(B-V)_{\rm AGN}$ was set to be from 0 to 1.3 with an interval of 0.1. Finally, we adopt the model from \citet{2014ApJ...784...83D} to approximate the dust emission from cold dust. The best-fit SED is shown in Figure~\ref{fig:SED}. The physical parameters from the best-fitting model are listed in Table~\ref{tab:SED_param}. Intriguingly, the extinction of the AGN with $E(B-V) \sim 1.0$ is significantly larger than that of the host galaxy with $E(B-V) \sim 0.2$, indicating a heavy obscuration of the nucleus possibly due to polar dust. The relatively high AGN fraction of $\sim$0.95 suggests that the IR emission of our target is mostly associated with vigorous AGN activity. However, the star formation rate (SFR) is comparable to that of starburst galaxies with a similar redshift \citep{2011ApJ...739L..40R}. A portion of the old stellar population is surprisingly higher than that of the young stellar population from the SED analysis, which hints star formation happened in the early universe. Overall, our spectral analysis suggests that ADFS-KMTDOG-102 is a massive starburst galaxy hosting a luminous AGN at its center, which dominates the IR emission.

\subsection{Black Hole Mass}
Previous studies demonstrated that the extreme infrared luminosity of the BlueDOG may originate from AGNs with black hole masses of $10^{8-10} M_\odot$ and super-Eddington ratio (e.g., \citealt{2022ApJ...941..195N, 2023ApJ...959L..14N}). Therefore, it is important to estimate the black hole mass of our target to understand its physical origin. First, by adopting the method in \citet{2011ApJS..194...45S} derived from the SDSS quasars, we calculated BH mass with the FWHM and extinction corrected luminosity of the broad H$\mathrm \alpha$ emission line. Secondly, based on Equation~(7) from \citet{2006ApJ...641..689V}, we additionally estimated BH mass with the FWHM of \ion{C}{4} and 1350${\mathrm{\AA}}$ monochromatic luminosity. Note that the H$\mathrm \alpha$ and 1350${\mathrm{\AA}}$ luminosities were corrected for the intrinsic extinction with $E(B-V)_{\mathrm{AGN}}=1.03$ estimated from the SED fitting (see details in the section~\ref{subsec:sed_fitting}). \citet{2023ApJ...954..156K} argued that the MIR-based BH mass estimator gives more reliable results for the BH mass measurements for dust-obscured quasars because the MIR luminosity is relatively free from the obscuration. To account for this systematic due to the extinction, we additionally estimated BH mass with the 4.6$\mu$m monochromatic luminosity and FWHM of broad H$\mathrm \alpha$, by employing the H$\mathrm{\alpha}$-4.6$\mu$m monochromatic luminosity relation (Table~3 in \citealt{2023ApJ...954..156K}).
Overall, the three estimates of the BH mass are consistent with each other within their respective uncertainties (Table~\ref{tab:SED_param}).

The uncertainties of BH mass are computed from the quadratic summation of measurement errors and the systematic uncertainties associated with each BH mass estimator. 

\begin{deluxetable}{lll}
    \tabletypesize{\footnotesize}
    \tablenum{2}
    \tablecaption{Photometry points from band-merged catalog\label{tab:photometry}}
    \tablehead{
        \colhead{Facility/Band} & \colhead{F$_\nu$} & \colhead{Reference}
    }
    \startdata
    Blanco/DECam g & 2.61$\pm$0.17 $\rm\mu$Jy &  {A18}\\
    Blanco/DECam r & 2.45$\pm$0.22 $\rm\mu$Jy&  {A18}\\
    Blanco/DECam i & 2.83$\pm$0.40 $\rm\mu$Jy&  {A18}\\
    Blanco/DECam z & 2.45$\pm$0.79 $\rm\mu$Jy&  {A18}\\
    Blanco/DECam Y & 5.29$\pm$2.64 $\rm\mu$Jy&  {A18}\\
    KMTNet/B &  2.55$\pm$0.20 $\rm\mu$Jy&  J16\\
    KMTNet/R &  2.46$\pm$0.15 $\rm\mu$Jy&  J16\\
    KMTNet/I &  2.43$\pm$0.21 $\rm\mu$Jy&  J16\\
    VISTA/VIRCAM H  & 34.35$\pm$8.96 $\rm\mu$Jy&  {M13}\\
    VISTA/VIRCAM Ks & 90.89$\pm$19.61 $\rm\mu$Jy&  {M13}\\
    WISE/W1 & 0.324$\pm$0.008 mJy &  {C14}\\
    WISE/W2 & 0.757$\pm$0.018 mJy &  {C14}\\
    WISE/W3 & 4.97$\pm$0.13 mJy &  {C14}\\
    WISE/W4 & 11.39$\pm$0.75 mJy &  {C14}\\
    Spitzer/IRAC Ch.1 & 0.379$\pm$0.003 mJy & {B16}\\
    Spitzer/MIPS 24$\mu$m & 9.65$\pm$0.02 mJy & {C11}\\
    Spitzer/MIPS 70$\mu$m & 29.0$\pm$6.0 mJy & {C11}\\
    AKARI/FIS 90$\mu$m & 35.0$\pm$5.4 mJy & {Private comm.}\\
    Herschel/PACS 100$\mu$m & 23.1$\pm$13.7 mJy & {O12}\\
    Herschel/SPIRE 250$\mu$m & 40.7 $\pm$2.9 mJy & {W14}\\
    Herschel/SPIRE 350$\mu$m & 35.8 $\pm$2.6 mJy & {W14}\\
    Herschel/SPIRE 500$\mu$m & 26.8 $\pm$2.9 mJy & {W14}\\
    \enddata
    \tablecomments{A18: \citet{2018ApJS..239...18A}; J16: \citet{2016JKAS...49..225J}; M13: \citet{2013Msngr.154...35M}; C14: \citet{2014yCat.2328....0C}; B16: \citet{2016ApJS..223....1B}; C11: \citet{2011MNRAS.411..373C}; O12: \citet{2012MNRAS.424.1614O}; W14: \citet{2014MNRAS.444.2870W}}
\end{deluxetable}

\begin{deluxetable}{hlhhch}
    \tabletypesize{\footnotesize}
    \tablenum{3}
    \tablecaption{Physical parameters from best-fit SED and SMBH mass measurement using available emission lines.\label{tab:SED_param}}
    \tablehead{
        \colhead{} & \colhead{Parameter} & \colhead{} & \colhead{} & \colhead{Value} & \colhead{}  
        } 
        \startdata
        \multicolumn{6}{c}{SED parameters}\\
        {} & $M_{\mathrm{stellar}}$ & {} & {} & (2.12$\pm 0.21$)$\times$10$^{12}$~M$_{\odot}$ & {}\\
        {} & SFR  & {} & {} &    278.2$\pm 51.6$~M$_{\odot}~yr^{-1}$ & {}\\
        {} & $L_{\mathrm{bol}}$ & {} & {} & (8.12$\pm 0.36$)$\times$ 10$^{13}$~L$_{\odot}$ & {}\\
        {} & $L_{\mathrm{bol,\,AGN}}$ & {} & {} & (7.13$\pm 0.36$)$\times$ 10$^{13}$~L$_{\odot}$ & {}\\
        {} & $f_{\mathrm{AGN}}$ & {} & {} & 0.95$\pm 0.001$ & {}\\
        {} & $E(B-V)_{\rm AGN}$ & {} & {} &  1.026$\pm 0.064$ & {}\\
        {} & $E(B-V)_{\rm stellar}$ & {} & {} & 0.20$\pm 0.001$ & {}\\
        {} & Age$_{\rm main}$$^{a}$ & {} & {} & 1.86$\pm 0.34$ [Gyr] & {} \\
        \hline
        \multicolumn{6}{c}{Measurement of SMBH mass$^{b}$}\\
        {} & log~($M_{\mathrm{BH,\, H\alpha}}$/$M_{\odot}$) & {} & {} & 10.15$\pm 0.25$ & {}\\
        {} & log~($M_{\mathrm{BH,\,C\,{IV}}}$/$M_{\odot}$) & {} & {} & 9.92$\pm 0.36$ & {}\\
        {} & log~($M_{\mathrm{BH,\,4.6\mu m}}$/$M_{\odot}$) & {} & {} & 9.75$\pm 0.22$ & {}\\
        \hline
        \multicolumn{6}{c}{Eddington ratio}\\
        {} & $\lambda_{\rm Edd,\,\mathrm{H\mathrm{\alpha}}}$ & {} & {} & 0.16$\pm 0.09$ & {}\\
        {} & $\lambda_{\rm Edd,\,\mathrm{C\,{IV}}}$ & {} & {} & 0.26$\pm 0.22$ & {}\\
        {} & $\lambda_{\rm Edd,\,\mathrm{4.6\mu m}}$ & {} & {} & 0.38$\pm 0.19$ & {}\\
        \enddata 
        \tablecomments{$^a$ Age of the main stellar population. $^b$Uncertainties of black hole mass are considered by the quadratic summation of measurement errors and the systematic scatter of each relation. 0.18 dex \citep{2011ApJS..194...45S}, 0.32 dex \citep{2006ApJ...641..689V}, and 0.20 dex \citep{2023ApJ...954..156K} are used to consider systematic scatters for H$\mathrm{\alpha}$, \ion{C}{4}, and $L_{4.6\mu m}$, respectively.}
\end{deluxetable}

\section{Discussion}\label{sec:discussion}
 
\subsection{Ly$\alpha$ escape fraction}\label{subsec:Lya_escape_frac}

Ly$\mathrm \alpha$ escape fraction ($f_{\rm esc}^{Ly\alpha}$) is widely used to measure the properties of galaxies in the context of star formation. Although the extinction is derived, estimating a Ly$\mathrm \alpha$ escape fraction ($f_{\rm esc}^{Ly\alpha}$) with extinction-corrected $L_{\mathrm{H\alpha}}$ and observed $L_{\mathrm{Ly\alpha}}$ remains challenging, since these values are also influenced by AGN activity. We try to disentangle the calculation of $f_{\rm esc}^{Ly\alpha}$ with SFR inferred Ly${\mathrm{\alpha}}$ luminosity compared to AGN component.

\subsubsection{Star-formation aspect}\label{subsubsec:Lya_escape_frac_SF}
If H$\mathrm \alpha$ emission line is not available, Ly$\mathrm \alpha$ can used as a star-formation tracer. Thus, the expected Ly${\mathrm{\alpha}}$ luminosity can be estimated from SFR \citep{2017ApJ...850..178A}. Here, we assume the case B recombination with $n_e\approx350$~cm$^{-3}$ and $T=10^4$~K in order to estimate an intrinsic Ly$\alpha$ luminosity from SED-derived SFR, which corresponds to the intrinsic Ly$\mathrm \alpha$/H$\mathrm \alpha$ ratio of 8.7. We calculate the intrinsic Ly$\alpha$ luminosity combining the intrinsic Ly$\mathrm \alpha$/H$\mathrm \alpha$ ratio of case B and below equation from \citet{1998ARA&A..36..189K}:
\begin{eqnarray}
    {\rm SFR}\,[M_\odot\,yr^{-1}] = 0.68\times 7.9\times 10^{-42} L_{\rm H\alpha}\,[{\rm erg\,s}^{-1}]\nonumber\\
    = 0.68\times 7.9\times 10^{-42}\times L_{\rm Ly\alpha}/8.7\,[{\rm erg\,s}^{-1}]
\end{eqnarray}\label{eq:L_Lya_intrinsic}
The 0.68 factor means a correction factor from Salpeter to IMF suggested by \citet{2001MNRAS.322..231K} \citep{2012ARA&A..50..531K}. Although Chabrier IMF is used in the SED fitting, we use the correction factor of Kroupa IMF as Chabrier IMF shows similar results to Kroupa IMF \citep{2012ARA&A..50..531K}. We derive $f_{\rm esc}^{\rm Ly\alpha}$ based on SFR-derived intrinsic Ly${\mathrm{\alpha}}$ luminosity with:
\begin{equation}
    f_{\rm esc,\,SF}^{\rm Ly\alpha}=L_{\rm Ly\alpha,\,obs}/L_{\rm Ly\alpha,\, int}
\end{equation}\label{eq:f_esc_SFR}
This yields $f_{\rm esc,\,SF}^{\rm Ly\alpha}$=17.6$\pm$3.3\%, which is consistent with the escape fraction observed in other LAEs with similar $E(B-V)_{SED}$ \citep{2014ApJ...791....3S, 2016MNRAS.458..449M} (Figure~\ref{fig:EBV_f_esc_Lya}). In Figure~\ref{fig:EBV_f_esc_Lya}, we adopt $E(B-V)_{\rm SED}$ as $E(B-V)_{\rm stellar}$ calculated from the SED fitting for $f_{\rm esc,\,SF}^{\rm Ly\alpha}$ because $f_{\rm esc,\,SF}^{\rm Ly\alpha}$ is derived from the SFR which represents the properties of the host galaxy.

\subsubsection{AGN aspect}\label{subsubsec:Lya_escape_frac_AGN}
\begin{figure}[t!]
    \includegraphics[scale=0.42, trim={2.cm 1cm 0 0cm}]{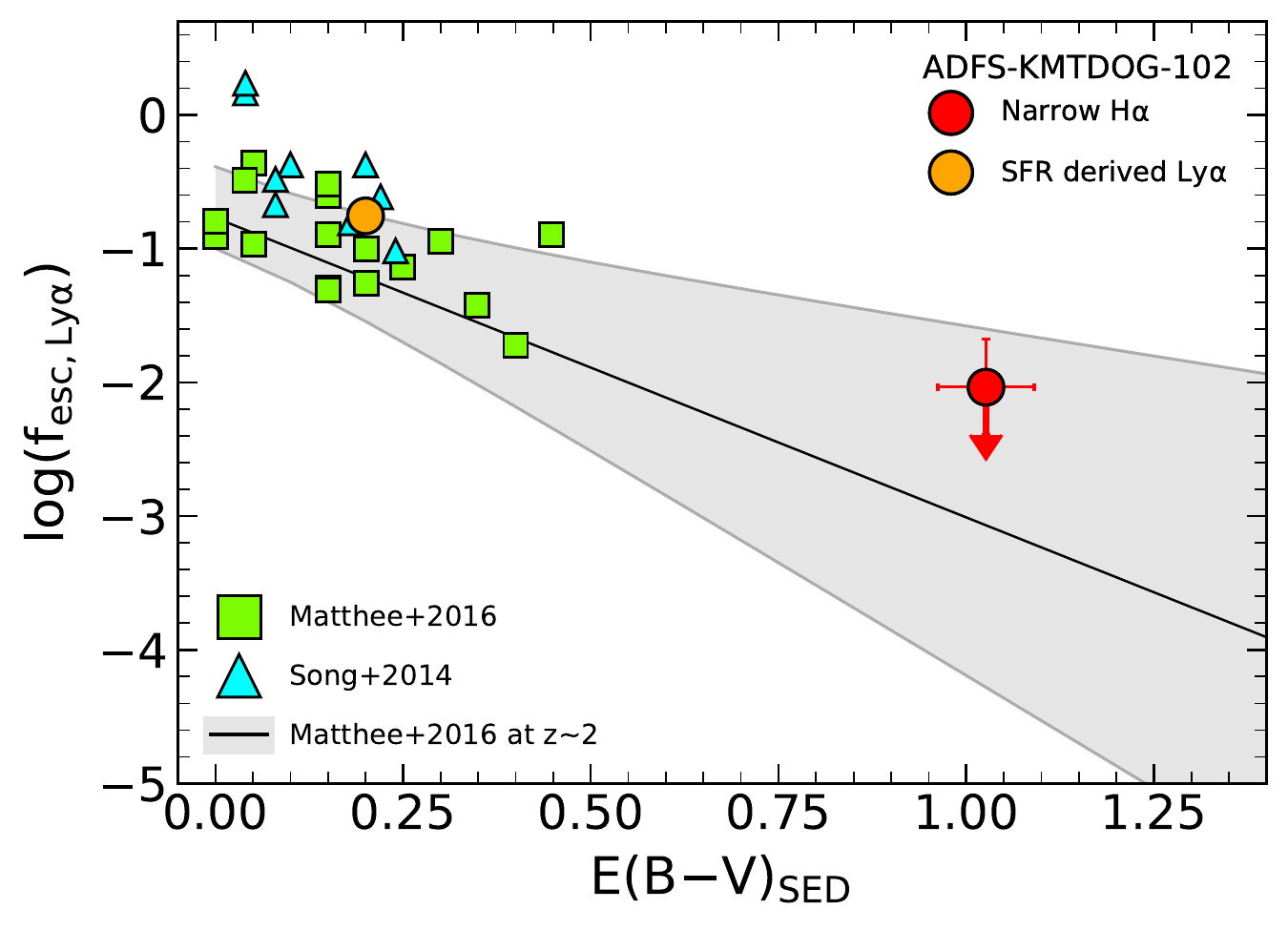}
    \caption{Ly$\mathrm{\alpha}$ escape fraction as a function of E(B$-$V) derived from SED fitting. We present the Ly$\mathrm{\alpha}$ escape fraction ($f_{\rm esc}^{\rm Ly\alpha}$) of the BlueDOG by dividing into SFR and AGN cases in orange and red circles, respectively. The upper limit of $f_{\rm esc}^{\rm Ly\alpha}$ in the AGN case is estimated with a narrow component of H$\mathrm{\alpha}$ assuming NLR condition. Also, the SFR-derived $f_{\rm esc}^{\rm Ly\alpha}$ of the BlueDOG is located near the relation of LAEs at similar redshift \citep{2014ApJ...791....3S, 2016MNRAS.458..449M}. The black line and gray shade show the fitted relation from \citet{2016MNRAS.458..449M}.  \label{fig:EBV_f_esc_Lya} } 
\end{figure}

The intrinsic Ly$\mathrm \alpha$/H$\mathrm \alpha$ ratio at the vicinity of BLR is highly uncertain due to collisional excitation. Thus, the case B recombination is unlikely in this case. However, \citet{1984PASP...96..393G} suggested intrinsic Ly$\mathrm \alpha$/H$\mathrm \alpha$ ratio varies from 11-16 in NLR conditions. Using the ratio, we calculate an upper limit of $f_{\rm esc,\,AGN}^{\rm Ly\alpha}$ only with a narrow H$\mathrm \alpha$ component. Then, the Ly$\mathrm \alpha$ escape fraction of AGN is given by:
\begin{equation}
    f_{\rm esc,\,AGN}^{\rm Ly\alpha}=L_{\rm Ly\alpha,\,obs}/(11L_{\rm H\alpha,\,narrow,\, corr})
\end{equation}\label{eq:f_esc_AGN}
L$_{\mathrm{H\alpha,\,narrow,\, corr}}$ is extinction corrected H$\mathrm \alpha$ luminosity of narrow component with $E(B-V)_{\rm AGN}$. We choose the intrinsic Ly$\mathrm \alpha$/H$\mathrm \alpha$ ratio of 11 in order to make estimated $f_{\rm esc,\,AGN}^{\rm Ly\alpha}$ as the upper limit. 
From the equation, we estimate $f_{\rm esc,\,AGN}^{\rm Ly\alpha}\leq0.9\pm 0.8\%$ at most, which means that Ly$\mathrm \alpha$ photons hardly escape from the system. Although the majority of Ly$\mathrm \alpha$ photons cannot pass through the system, the observed Ly$\mathrm \alpha$ luminosity remains exceptionally high. The calculated $f_{\rm esc,\,AGN}^{\rm Ly\alpha}$ is still located in the 1-sigma uncertainty of a relation suggested from \citet{2016MNRAS.458..449M} at \textit{z}$\sim$2 (Figure~\ref{fig:EBV_f_esc_Lya}). Therefore, we suggest that the SFR-derived Ly$\mathrm \alpha$ escape fraction might be overestimated compared to the realistic value, if an AGN is present.

\subsection{SED modeling without polar dust of AGN}\label{subsec:discuss_SED_wo_polar_dust}
In the absence of polar dust emission, as modeled with torus framework proposed by \citet{2006MNRAS.366..767F}, dust emission contributes more significantly to the FIR SED than the AGN SED. It returns SFR$\sim$8226~M$_{\odot}$~yr$^{-1}$ and $E(B-V)_{\rm stellar}=0.31$ to compensate for the absence of AGN dust contribution by increasing SFR and decreasing extinction without polar dust. However, such an extreme starburst with SFR of more than a thousand solar masses per year is unlikely in the relatively unobscured environment indicated by $E(B-V)_{\rm stellar}$. \citet{2020ApJ...897..112A} discussed the case of unobscured extreme starburst for the origin of blue-excess emission in blue hot dust-obscured galaxies (BHDs). They insisted that SFR of $\gtrsim1000$~M$_{\mathrm{\odot}}$~yr$^{-1}$ is unlikely in unobscured star formation. \citet{2018ApJ...866...92L} presented SED modeling with polar dust in type 1 AGNs. They suggested that HotDOGs have a large optical depth of polar dust ($\tau_V$) $\sim$ 5$-$20 and ERQs have intermediate $\tau_V\sim$3.0 ($A_V=1.086\tau_V$). Note that ERQs show SED similar to BlueDOGs and BHDs. 

Thus, a large amount of polar dust (i.e., large $E(B-V)_{\rm AGN}$) is more plausible than the unobscured starburst, which shows probable physical conditions in SFR and extinction. Although we use more plausible models to derive SED properties, a caveat remains that most UV emission is dominated by stellar components rather than AGN SED, which contrasts with the origin of UV emission lines. These UV emissions are discussed in detail in section~\ref{subsec:implication_from_line_ratio}.

\subsection{SMBH mass and accretion rate}\label{subsec:SMBH_efficiency}
The SMBH mass estimates derived from H$\mathrm \alpha$, \ion{C}{4}, and MIR-based are consistent with one another, within the margin of uncertainty (Table~\ref{tab:SED_param}). 
All three measurements lie above the galaxy$-$SMBH mass relation for local broad-line AGNs from \citet{2015ApJ...813...82R}, with a specific offset of 1.30 dex in M$_{\rm BH,\,H\alpha}$. This may suggest that the SMBHs of massive high-z galaxies rapidly accumulated their mass even in the early universe.
Recent studies have discovered that SMBHs are already formed at \textit{z}$\sim$3-7 (\citealt{2015ApJ...806..109J}; \citealt{2015Natur.518..512W}; \citealt{2021ApJ...914...36I}), supporting the feasibility of the stellar mass and SMBH mass observed in our BlueDOG.

The proxy of accretion rate, the Eddington ratio ($\lambda_{\rm Edd}$), can be defined as:
\begin{equation}\label{eq:Eddington_ratio}
    \lambda_{\rm Edd}=\frac{L_{\rm bol,\,AGN}}{L_{\rm Edd}},
\end{equation}
where $L_{\rm Edd}=3.28\times10^4 \left(\frac{M_{\rm BH}}{M_{\odot}}\right) L_\odot$, Eddington luminosity. Using the equations~\ref{eq:Eddington_ratio} and $L_{\rm bol}$ from SED fitting, we derive the Eddington ratio for each method. The Eddington ratios are $\lambda_{\rm Edd}$=0.16$\pm 0.09$, 0.26$\pm 0.22$, and 0.38$\pm 0.19$ for H$\mathrm \alpha$, \ion{C}{4}, and $L_{4.6\mu m}$-based SMBH mass, respectively. 
\citet{2022ApJ...941..195N} estimated SMBH masses and Eddington ratios from \ion{C}{4} emission of four BluDOGs, where the most massive SMBH mass is 1 dex less than that of our BlueDOG but with super Eddington ratio ($\lambda_{\rm Edd}\gtrsim1$). Our BlueDOG shows $\sim$10 times less Eddington ratio compared to \citet{2022ApJ...941..195N}. This could be caused by the massive SMBH of the BlueDOG, which has $\sim$2 dex larger SMBH mass. The red quasars from \citet{2012MNRAS.427.2275B} also show a similar Eddington ratio and $A_{V}$ with $\sim$2-6 compared to the BlueDOG, which implies that they might be experiencing a similar evolutionary phase.

\subsection{Similarity to Little Red Dots discovered by JWST}
James Webb Space Telescope (JWST) has recently identified Little Red Dots (LRDs), which host unexpectedly massive BHs at such early epoch of \textit{z}$\sim$5-8 with ``V-shape of SED'' at rest-frame of UV and optical. These LRDs also show a SED shape similar to that of BlueDOGs and BHDs. However, the origin of the blue-excess UV emission at LRDs is still controversial between scattered light from AGN and extended star-formation, although the broad H$\rm{\alpha}$ emission line and the power-law MIR slope are plausible to be dominated by AGN (e.g.,  \citealt{2023arXiv231203065K}; \citealt{2023arXiv230607320L}; \citealt{2024arXiv240807745B}; \citealt{2024ApJ...963..128B}; \citealt{2024ApJ...964...39G}; \citealt{2024arXiv240403576K}; \citealt{2024ApJ...963..129M}; \citealt{2024arXiv241103424S}).

\begin{figure*}[t!]
    \centering
    \includegraphics[height=5.8cm, width=8.5cm, trim={0 0.5cm 0 0}]{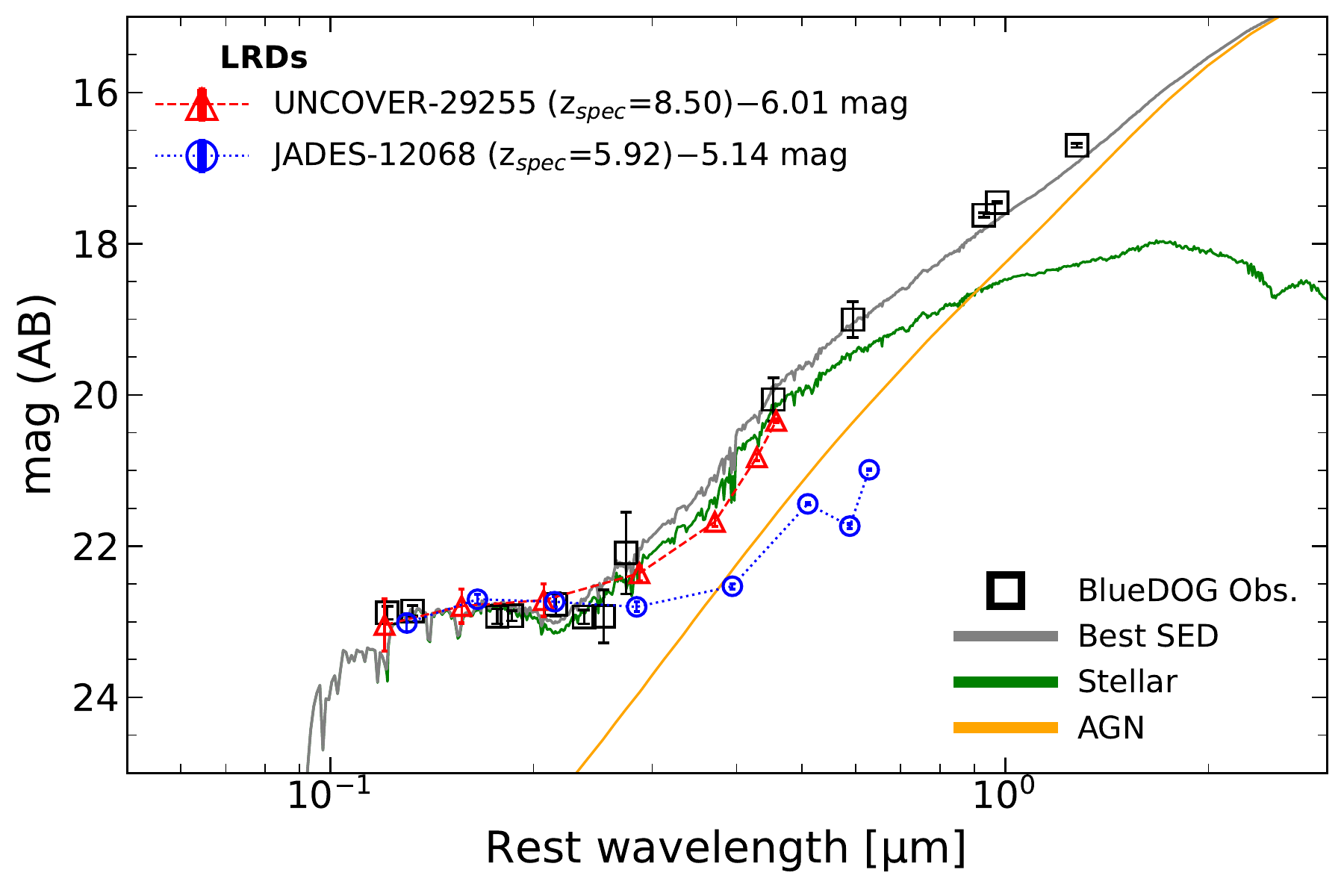}
    \includegraphics[height=5.8cm, width=8.5cm, trim={0 0.5cm 0 0}]{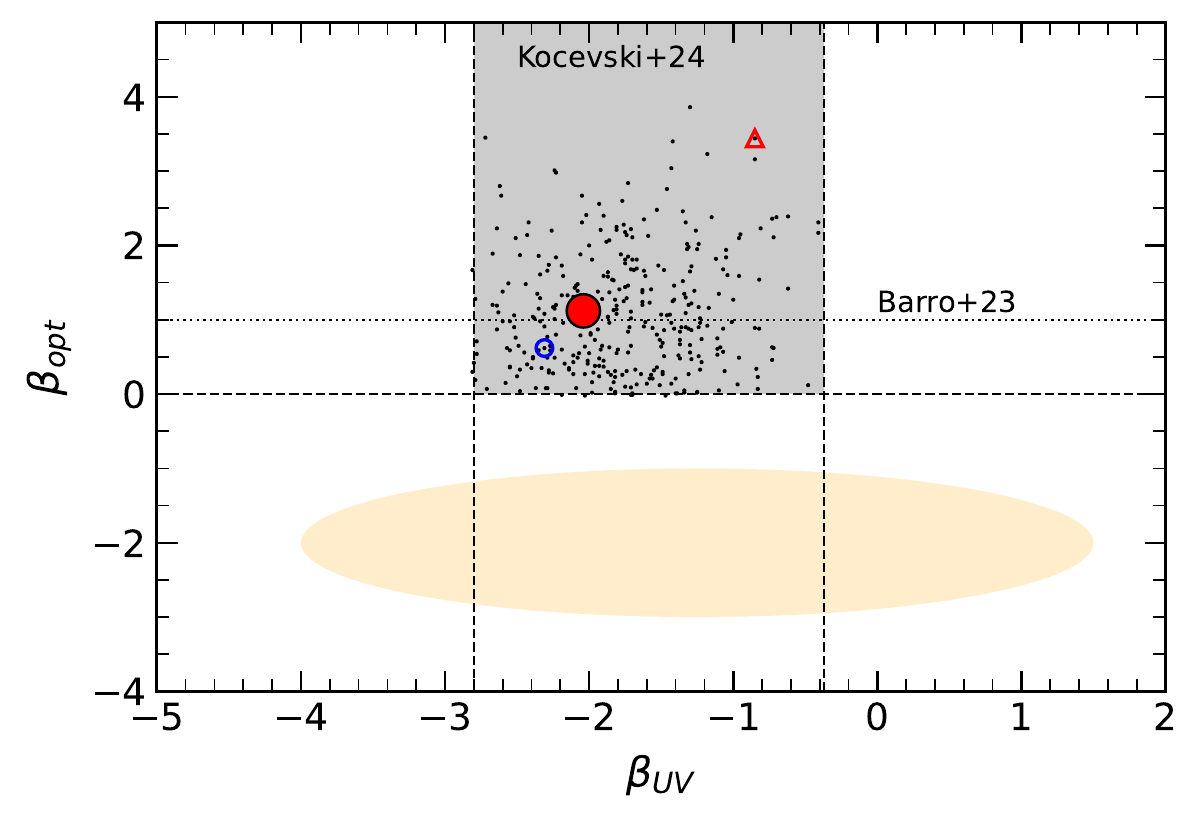}
    \caption{(Left) SEDs of the BlueDOG and LRDs are overplotted. The observed mags of the BlueDOG are shown in black squares. Solid lines are best fit SED for the BlueDOG (gray: total SED; green: stellar; orange: AGN). Two LRDs shown as triangle and circle represent each type of LRDs from \citet{2024arXiv240403576K}. Their mags are normalized to the BlueDOG at 1500$\rm{\AA}$ by adding normalization mag written at each marker of LRDs. We use the photometric catalog for LRDs from ASTRODEEP-JWST \citep{2024arXiv240900169M}. Although we used the fluxes of UNCOVER-29255, which are highly lensed by a foreground galaxy cluster, Abell 2744, it is not critical on the normalized SED since flux boosting by gravitational lensing is independent of wavelength. (Right) Selection criteria for LRDs from \citet{2024arXiv240403576K} and \citet{2024ApJ...963..128B} are shown as gray shaded region plus dashed lines and dotted line, respectively. We also present the measured UV and optical slopes of LRDs from \citet{2024arXiv240403576K} in the black dots. The filled red circle represents the measurement of the BlueDOG, suggesting that the BlueDOG can also be classified as an LRD. The open triangle and circle are the two LRDs shown in the left panel. The yellow ellipse stands for the distribution of slopes in typical galaxies observed with JWST (See the detail in Figure~3 of \citealt{2024arXiv240403576K}).
    \label{fig:SED_KMTDOG102_LRD}}
\end{figure*}

In Figure~\ref{fig:SED_KMTDOG102_LRD}, we measure UV and optical slopes of the BlueDOG at blueward and redward from rest-frame 3645$\rm\AA$ respectively, which is described in \citet{2024arXiv240403576K}. The UV and optical slopes of the BlueDOG meet both the LRD criteria suggested by \citet{2024arXiv240403576K} and \citet{2024ApJ...963..128B} with slopes of $\beta_{\rm UV}$=-2.04$\pm$0.10 and $\beta_{\rm opt}$=1.12$\pm$0.13. Among 43 redshift-confirmed LRDs from \citet{2024arXiv240403576K}, combining spectroscopic redshifts from ASTRODEEP-JWST \citep{2024arXiv240900169M}, we identified two representative types of SEDs. Although both exhibit similar slope and shape at rest-frame UV, one shows a steeper slope (e.g., UNCOVER-29255), while the other has a gentle slope (e.g., JADES-12068) at rest-frame optical to NIR, as shown in Figure~\ref{fig:SED_KMTDOG102_LRD}. However, the optical slope of UNCOVER-29255, presented in the right panel of Figure~\ref{fig:SED_KMTDOG102_LRD}, appears overestimated due to the lack of MIR photometric points. Consequently, the normalized SED shape of UNCOVER-29255 becomes more similar to that of the BlueDOG. If we assume that the similarity in SEDs between the BlueDOG and UNCOVER-29255 arises from the same physical mechanism (i.e., same SED parameters), the primary difference between the BlueDOG and UNCOVER-29255 may lie in their masses (e.g., stellar mass and dust mass), as the magnitudes of LRDs were scaled for comparison. What drives the different optical to NIR slopes between the two types of LRDs? The first case assumes that the two types of LRDs have different old stellar population masses with a similarly high AGN fraction. The second case suggests that the two LRDs have different AGN fractions while maintaining similar old stellar population masses. However, the first case is unlikely for UNCOVER-29255 because the cosmic age of 0.58\,Gyrs at \textit{z}=8.5 is not sufficient for stars to evolve into an old stellar population.

Spectroscopic analysis of LRDs reveals properties similar to those of the BlueDOG. \citet{2024ApJ...963..129M} presented broad H$\rm{\alpha}$-detected LRDs have SMBH mass of $10^{7-8}\,M_\odot$ and Eddington ratio of $\lambda_{\rm Edd}\sim0.07-0.4$ (typically 0.16). The BlueDOG's Eddington ratio of $\lambda_{\rm Edd}=0.16\pm0.09$, similar to that of LRDs, suggests that both may be undergoing a similar accretion state at their central SMBH. Regardless of those similarities, both the BlueDOG and LRDs host a surprisingly massive BH at their respective epochs.

\subsection{Scenario for Origin of Blue-excess UV Emission}\label{subsec:implication_from_line_ratio}
\subsubsection{UV Emission lines}\label{subsubsec:origin_of_emlines}
The observed line ratio of Ly$\mathrm \alpha$/\ion{N}{5}$_{1240}$ is 2.04 from the BlueDOG, which is smaller than the line ratio measured from Lyman Break Galaxies (LBG; Ly$\mathrm \alpha$/\ion{N}{5}$_{1240}\sim15$) at \textit{z}$\sim$3 \citep{2003ApJ...588...65S}, faint narrow-lined AGNs at \textit{z}$\sim$2$-$3 (\citealt{2002ApJ...576..653S}; \citealt{2011ApJ...733...31H}; Ly$\mathrm \alpha$/\ion{N}{5}$_{1240}\sim11.86$), and a star-forming galaxy at \textit{z}=5.34 (\citealt{1998ApJ...498L..93D}; Ly$\mathrm \alpha$/\ion{N}{5}$_{1240}\sim100$). 
On the other hand, the composite spectra from high-z broad-lined AGNs at \textit{z}$\sim$3 show a larger \ion{N}{5}$_{1240}$ flux compared to the composite of narrow-lined AGNs, although an exact value is not provided in the literature (See Fig~1 in \citealt{2002ApJ...576..653S}). It suggests that Ly$\mathrm \alpha$/\ion{N}{5}$_{1240}$ from high-z broad-lined AGNs should be lower than that of narrow-lined AGNs at similar redshift. It is natural that type 1 AGNs have a significantly low Ly$\mathrm \alpha$/\ion{N}{5}$_{1240}$ ratio, because the N$^{3+}$ ion has quite high ionization energy of 77~eV to be ionized to N$^{4+}$.
\citet{2017MNRAS.464.3431H} also showed that core extremely red quasars (core ERQ) have small Ly$\mathrm \alpha$/\ion{N}{5}$_{1240}$ ratio at \textit{z}$\sim$2$-$3.4. Similarly, \citet{2020A&A...634A.116V} presented 21 core ERQs which exhibit median Ly$\mathrm \alpha$/\ion{N}{5}$_{1240}\sim$0.94$\pm 0.15$ at \textit{z}$\sim$2$-$3. 
The Ly$\mathrm \alpha$/\ion{N}{5}$_{1240}$ ratio from our BlueDOG, therefore, is consistent with the line ratio of obscured type 1 AGNs, which suggests the UV emission lines of the BlueDOG might originate in the central AGN rather than the starburst region. For core ERQs, \citet{2020A&A...634A.116V} suggested that we are observing only the outer part of BLR with unseen inner BLR because a high fraction of core ERQs are lying at an intermediate orientation between type 1 and type 2. This scenario is partially consistent with our results, however, the coexistence of extremely broad H$\mathrm \alpha$ and relatively narrow UV emission lines is unexpected in our BlueDOG (FWHM in Table~\ref{tab:line_fitting_param}). 
This suggests that while we observe the inner BLR at least via the broad H$\mathrm \alpha$, we can not observe directly the inner BLR with Ly$\mathrm \alpha$ even other UV emission lines for the BlueDOG due to heavy obscuration in the direction of the AGN. 

\citet{1996ApJ...464..158H} estimated that the scattered light from broad \ion{N}{5} in non-BAL QSOs contributes up to 31\% of the total flux. Although they concluded that scattering does not significantly enhance the strength of \ion{N}{5} and \ion{C}{4}, the contribution of scattered light in these lines cannot be negligible. This suggests that the scattered light from central AGN in broad resonant lines with high ionization energy is a plausible explanation for the lower Ly$\rm\alpha$/\ion{N}{5} ratio observed in our source. 

\subsubsection{Contribution of emission lines to broadband photometry}\label{subsubsec:synthetic_photo}
We investigate the contribution of strong and broad emission lines to broadband photometry to quantify the amount of emission lines responsible for the blue excess emission. Using the emission line fitting result in section~\ref{subsec:emission_line_fitting}, we perform synthetic broadband photometry with fitted Gaussians rather than the optical spectra to reduce contamination from noise using \textit{Pyphot}. Especially, we focus on the $B$-band, which shows blue excess in the SED and covers the entire Ly$\mathrm \alpha$ and \ion{N}{5} emission lines. Assuming the total flux in the B-band as $F_{\rm syn,\,total}$=$F_{\rm syn,\,continuum}$+$F_{\rm syn,\,Ly\alpha}$+$F_{\rm syn,\,NV}$, we compare each relative ratio. The ratios of each component to total flux are $F_{\rm syn,\,Ly\alpha}$/$F_{\rm syn,\,total}=0.40$, $F_{\rm syn,\,NV}$/$F_{\rm syn,\,total}=0.21$, and $F_{\rm syn,\,continuum}$/$F_{\rm syn,\,total}=0.40$. The contributing Ly$\mathrm \alpha$ flux ratio is comparable to the ratio of continuum flux at the $B$-band. Furthermore, the sum of Ly$\mathrm \alpha$ and \ion{N}{5} emission lines accounts for about 60\% of total flux. This suggests the emission lines from the central AGN are more contributing to the broadband flux than continuum emission in the B-band flux of the BlueDOG. \citet{2023arXiv230100017M} showed that broad emission lines from optical spectra of hot DOGs can boost the broadband flux of JWST/NIRCam photometry about 25$-$80\%. For each emission line, they concluded the [\ion{O}{3}]+H$\mathrm \beta$ enhance the broadband flux by 30\%, and H$\mathrm \alpha$+[\ion{N}{2}] by 60\%. \citet{2022ApJ...941..195N} also showed that strong emission lines cause blue-excess emission for two BluDOGs in HSC $g$ and $r$ bands. Thus, these results suggest that we should carefully consider the contribution of strong broad emission lines, as they can significantly affect broadband photometry when the wavelength of the emission line falls within the broadband filter range.
However, the broadband flux boosting of the emission line does not seem significant on other broadbands since \ion{C}{4} contributes to broadband $R$-band flux about $\sim$1\% in the UV spectrum of the BlueDOG.

\begin{figure*}[t!]
    \centering
    \includegraphics[scale=0.7, trim={1.0cm 0.5cm 0 0cm}]{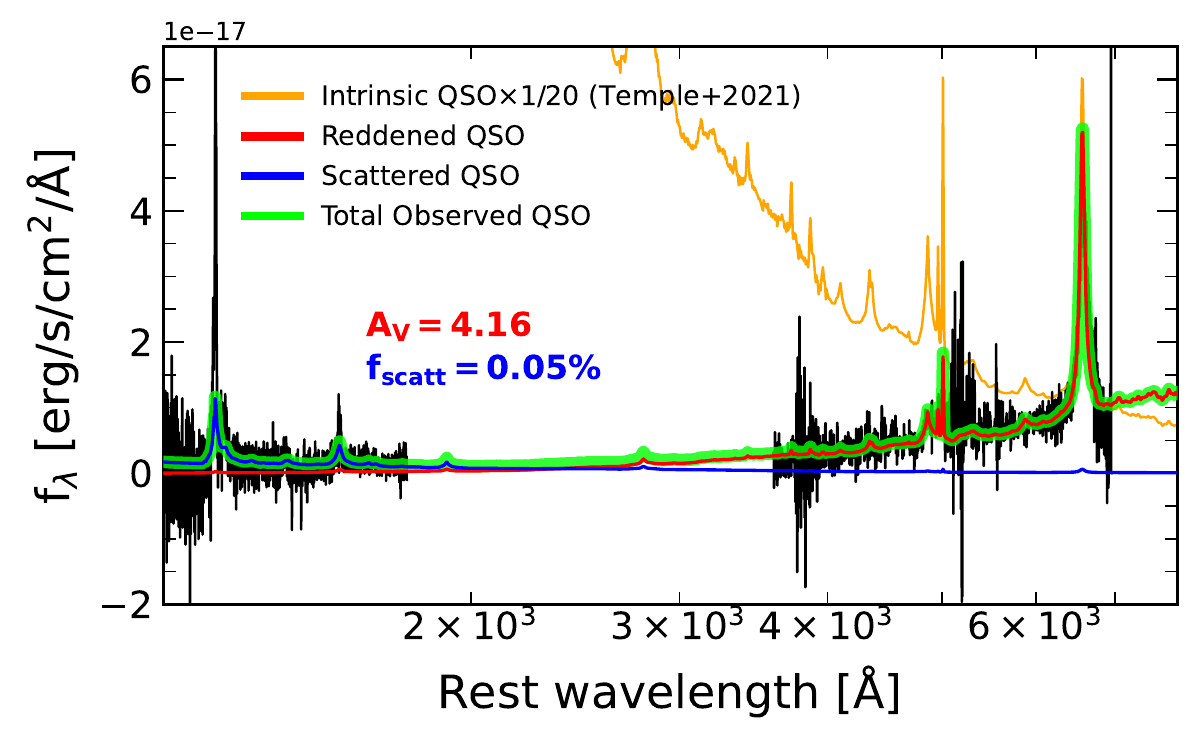}
    \caption{Modeling of the scattered light scenario using an intrinsic QSO template from \citet{2021MNRAS.508..737T}. The observed spectra are shown from \textit{GMOS} and \textit{FLAMINGOS-2} in black. For comparison, the intrinsic QSO SED is displayed in orange, scaled down by a factor of 1/20 for visualization. The red curve represents the reddened SED attenuated using $E(B-V)_{AGN}$ value derived from the CIGALE SED fitting. The blue SED shows a scattered component, modeled with a scattering fraction of 0.05\%. The green line represents the total modeled QSO SED, comprising the reddened and scattered components. \label{fig:BlueDOG_scattered_light_model}}
\end{figure*}

\subsubsection{UV Continuum}\label{subsubsec:origin_of_UV_continuum}
From the SED fitting with rigid SPIRE FIR detections described in Section~\ref{subsec:sed_fitting}, we discuss the plausible origin of the UV continuum in this section. In SED analysis, we derive that the best burst age is 12\,Myrs. We also estimated the specific star-forming rate (sSFR) within 10\,Myrs and 100\,Myrs in addition to the whole averaged sSFR. The sSFR, sSFR$_{\rm 10Myrs}$, and sSFR$_{\rm 100Myrs}$ are 0.132$\pm$0.028\,Gyr$^{-1}$, 0.168$\pm$0.030\,Gyr$^{-1}$, and 0.03$\pm$0.03\,Gyr$^{-1}$, respectively. Although the sSFR within 10\,Myrs is comparable to the average sSFR in whole, the sSFR within 100\,Myrs is approximately 6 times smaller than the sSFR$_{\rm 10Myrs}$. This indicates that the blue-excess UV continuum can be explained solely by stellar emission. Similarly, recent studies (e.g., \citealt{2024arXiv240807745B,2024arXiv241006257M,2024arXiv241103424S,2024arXiv240302304W}) have discovered a Balmer break in LRDs with JWST, which implies recent star formation within a few hundred Myrs. Thus, a recent starburst rather than a leaked AGN continuum could be responsible for the blue UV continuum in both the BlueDOG and LRDs. 

In contrast to the star-formation scenario, a scattered light model from a central AGN has been proposed to explain the excess of UV continuum (\citealt{2022ApJ...934..101A} for BHD; \citealt{2024ApJ...964...39G} for LRDs). \citet{2022ApJ...934..101A} reported a linear polarization of about 11\% in the UV flux of a BHD through polarimetric observation, suggesting that scattered AGN light contributes significantly to the observed blue excess. Similarly, \citet{2024ApJ...964...39G} modeled the contribution of scattered AGN light in LRDs and found that a combination of a reddened component with $A_V=2.7$ and the scattered component with f$_{\rm{scatt}}$=2.5\% successfully reproduces the ``V-shape'' of SED. Following the methodology of \citet{2024ApJ...964...39G}, we estimate the scattered light fraction required to explain the UV continuum excess in our BlueDOG. In the scattered light modeling, we fix only the $E(B-V)$ to $E(B-V)_{\rm AGN}$=1.03. Then, an intrinsic QSO luminosity and a scattered fraction are treated as free parameters. With the fixed $E(B-V)$, we find that the best scattered light model, based on the QSO template from \citet{2021MNRAS.508..737T}, requires a normalized intrinsic QSO luminosity of $\log (L_{3000}/{\rm erg\,s^{-1}})=47.7$ and a scattered fraction as low as 0.05\% to reproduce the observed UV continuum level (Figure~\ref{fig:BlueDOG_scattered_light_model}).
This implies that even a modest amount of scattered light could plausibly produce the observed UV continuum excess. However, this scattered light modeling still has important caveats. The bolometric correction (BC) derived by \citet{2006ApJS..166..470R} was obtained from blue quasars, and thus yields significantly different ratios when compared to the $L_{\mathrm{IR}}$, $L_{\mathrm{opt}}$, and $L_{\mathrm{bol}}$ of the BlueDOG. For example, \citealt{2006ApJS..166..470R} report median values of $L_{\mathrm{IR}}$/$L_{\mathrm{bol}}$=0.52 and $L_{\mathrm{opt}}$/$L_{\mathrm{bol}}$=0.28, whereas the BlueDOG shows $L_{\mathrm{IR}}$/$L_{\mathrm{bol}}$=0.97 and $L_{\mathrm{opt}}$/$L_{\mathrm{bol}}$=0.03. This discrepancy suggests that the BlueDOG has intrinsically different SED properties compared to the BC samples, making it difficult to directly apply previously established BC values. Moreover, the fact that the $L_{3000}$ adopted in the scattered light modeling exceeds the SED-integrated $L_{\mathrm{bol}}$ indicates that a fully scattered-light origin for the UV continuum is unlikely to explain the observations on its own. Therefore, we keep both the star-formation and scattered-light scenarios open in our interpretations. We note that the ambiguity between the star-formation and AGN-scattered light scenarios has a negligible impact on the derived physical parameters, particularly the stellar mass. Even under the extreme assumption that all UV continuum arises from AGN scattering, the contribution from young stellar populations becomes insignificant, and the stellar mass estimate still remains dominated by the old stellar component, as described in Section~\ref{subsec:sed_fitting}. 

\subsection{Metal Abundance of Circumnuclear Gas}\label{subsec:metallicity}
\begin{figure}[t!]
    \includegraphics[scale=0.37, trim={1.5cm 1cm 0 0cm}]{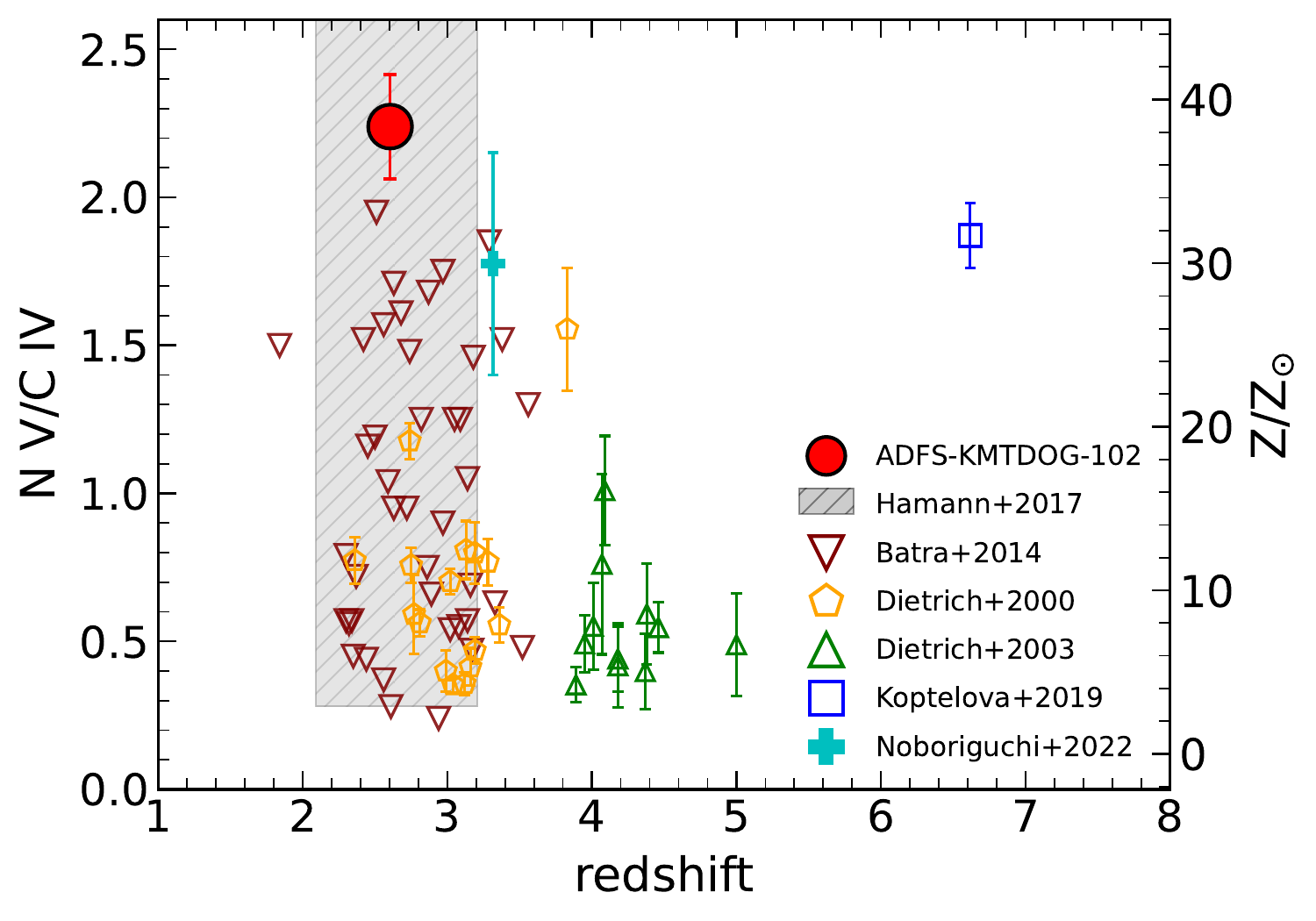}
    \caption{\ion{N}{5}/\ion{C}{4} line ratio as a function of redshift. Our measurement is shown in a red filled circle. Line ratios from other literature are plotted with various symbols. (gray shade) core ERQ samples \citep{2017MNRAS.464.3431H}. (inverted triangle) N-loud quasars at \textit{z}$\sim$2-3.5 \citep{2014MNRAS.439..771B}. (green triangle and pentagon) high redshift quasars \citep{2000A&A...354...17D, 2003ApJ...589..722D}. (blue square) an actively accreting quasar at \textit{z}=6.62 \citep{2019ApJ...882..144K}. (cyan plus) DOG with blue-excess emission \citep{2022ApJ...941..195N}. The right side of the y-axis means the corresponding metallicity to the \ion{N}{5}/\ion{C}{4} ratio. \label{fig:NV_CIV_metallicity}}
\end{figure}

The strong \ion{N}{5}$_{1240}$ emission line has been considered as supersolar metallicity with $Z>10Z_{\odot}$ in the BLR (\citealt{1993ApJ...418...11H}; \citealt{1999ARA&A..37..487H}; \citealt{2003ApJ...589..722D}; \citealt{2006A&A...447..157N}; \citealt{2014MNRAS.439..771B}; \citealt{2019ApJ...882..144K}; \citealt{2020A&A...634A.116V}). \citet{2017MNRAS.464.3431H} reported the \ion{N}{5}$_{1240}$/\ion{C}{4}$_{1549}$ ratio with a median value of 1.44 in 97 core ERQ samples. Similarly, \citet{2014MNRAS.439..771B} found that 41 N-loud quasars at \textit{z}$\sim$2.0-3.5 exhibit a median \ion{N}{5}$_{1240}$/\ion{C}{4}$_{1549}$ ratio of 0.95, corresponding to an estimated median metallicity of 5.5$Z_\odot$. \citet{2020A&A...634A.116V} suggested the UV lines in core ERQ originate from a metal-rich BLR. Similarly, we suggest that the BlueDOG have a metal-rich BLR, as its \ion{N}{5}$_{1240}$/\ion{C}{4}$_{1549}$ ratio of 2.22 is relatively high compared to other populations (Figure~\ref{fig:NV_CIV_metallicity}). The \ion{N}{5}$_{1240}$/\ion{C}{4}$_{1549}$ line ratio of the BlueDOG corresponds to $Z\sim37Z_{\odot}$, as estimated by extrapolating the results of \citet{2006A&A...447..157N} who analyzed the relationship between the line ratio and metallicity using a large sample of 5,344 SDSS quasars. Although \citet{2002ApJ...567L..19S} suggested the \ion{N}{5}$_{1240}$/\ion{C}{4}$_{1549}$ ratio as a metallicity tracer follows the line ratio-metallicity relation for only luminous case, the metallicity relation can be applied to our study since the BlueDOG has enormous luminosity where it is located at the end of luminous-regime in their Figure~2. 

\citet{2011A&A...527A.100M} found that more massive SMBHs have larger \ion{N}{5}$_{1240}$/\ion{C}{4}$_{1549}$ ratio with SDSS quasars. They suggested that AGB stars at the post-starburst phase can enrich the central region with nitrogen on a timescale of $\sim10^8$ years. On the other hand, carbon is mainly made by low-mass stars on a timescale of $\sim10^{9-10}$ years. Thus, the high metallicity of the BlueDOG originates from nitrogen overabundance in the central region, rather than high ISM metallicity throughout the entire host galaxy (See also \citealt{2019ApJ...882..144K}). This suggests that the star-formation history of the BlueDOG may be linked to the enrichment timescale of individual elements, as indicated by the SED fitting, which estimates the age of the main stellar population to be approximately 1.9\,Gyr (Table~\ref{tab:SED_param}). This age$_{\rm main}$ is consistent with the enriching timescales of each metal, which drive nitrogen overabundance and carbon deficiency. Also, there is another scenario for nitrogen enrichment in the aspect of stellar evolution. Such nitrogen enrichment is commonly observed in globular clusters (e.g., \citealt{2004ARA&A..42..385G}; \citealt{2005A&A...433..597C}; \citealt{2024ApJ...966...92S}). \citet{2024arXiv240408884R} discovered spectral features of Wolf-Rayet (WR) stars in the Sunburst Arc, a strongly lensed galaxy at \textit{z}=2.37. They presented low oxygen abundance and extreme nitrogen enrichment in a proto-globular cluster that hosts WR stars. Recent studies demonstrated nitrogen enrichment in high-redshift galaxies, including GN-z11 (\citealt{2023A&A...677A..88B}; \citealt{2023MNRAS.523.3516C}; \citealt{2023ApJ...959..100I}; \citealt{2024ApJ...966...92S}; \citealt{2024Natur.627...59M}; \citealt{2024MNRAS.529.3301T, 2024arXiv240719009T}). \citet{2023ApJ...959..100I} attributed this metal abundance to the presence of WR stars and supermassive stars, and tidal disruption events (TDE). Furthermore, \citet{2024MNRAS.529.3301T, 2024arXiv240719009T} identified dense, nitrogen-enriched gas at z$>$6 \ion{C}{4} emitters, suggesting a potential link between dense environment and nitrogen enrichment.

\begin{figure*}[t!]
    \includegraphics[width=19cm, height=10cm, trim={3.5cm 1cm 0 0cm}]{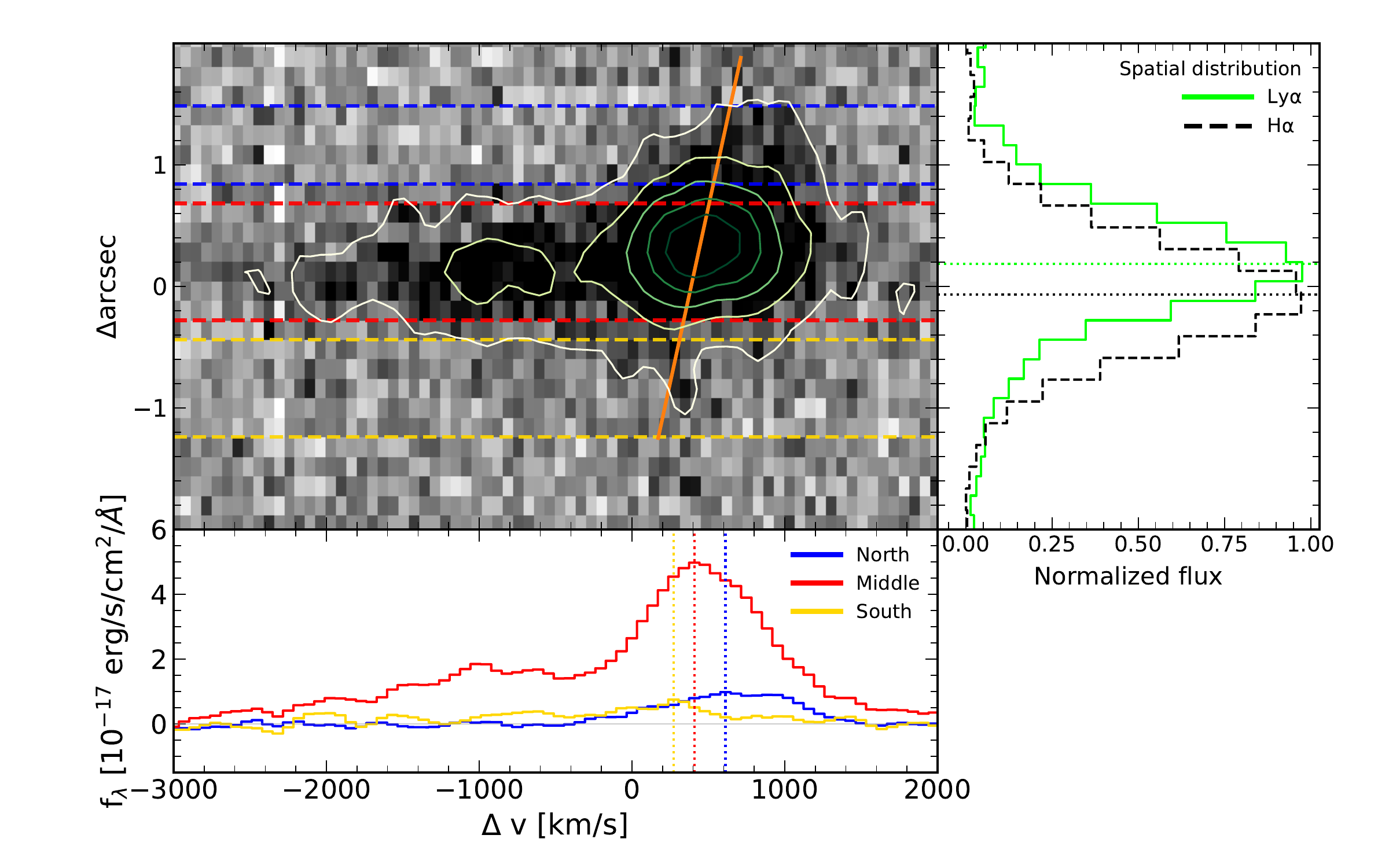}
    \caption{Ly$\mathrm{\alpha}$ spatial distribution analysis. (top left) 2D spectrum of Ly$\mathrm{\alpha}$ emission is shown as gray scale and contours. $\Delta arcsec$ of 0 corresponds to the position of the BlueDOG. We divided the spatial region of the Ly$\mathrm{\alpha}$ into color-coded dashed lines with blue, red, and yellow. The orange solid line is plotted with the peak position of each divided spatial region. (top right) Spatial profile comparison with Ly$\mathrm{\alpha}$ and H$\mathrm{\alpha}$. Dotted lines are the mean value of Gaussian fitting for each line. H$\mathrm{\alpha}$ has a shift of -0.07\,arcsec, which is negligible. However, Ly$\mathrm{\alpha}$ shows peak shift of 0.19\,arcsec. (bottom) Extracted spectrum from bounded regions by dashed lines in the top left panel. The vertical dotted lines represent the peak position of the extracted spectrum from three regions. The colors correspond to the colors from the top left panel.\label{fig:Lya_spatial_distribution}} 
\end{figure*}

\subsection{Nature of Ly$\rm{\alpha}$ emission}\label{subsec:lya_nature}
The profile of Ly$\rm \alpha$ emission shown in Figure~\ref{fig:spectrum} suggests that Ly$\rm \alpha$ photons escape through a scattering process in an expanding medium (See Figure~1 from \citealt{2014ApJ...793..114Y} for schematic view of Ly$\rm \alpha$ scattering process), since Ly$\rm \alpha$, a resonant line, can be scattered by hydrogen. \citet{2015A&A...578A...7V} suggested that sources with larger Ly$\rm \alpha$ peak separation have higher hydrogen column densities ($n_{\rm H I}$), corresponding to at least $\log\ (n_{\rm H I}) > 21$.  Thus, the scattering medium is expected to be hydrogen-rich as shown in X-ray observation of DOGs \citep{2024MNRAS.531..830K}. An expanding medium is a plausible explanation, since outflows are commonly observed in hyperluminous obscured AGNs (e.g., \citealt{2018ApJ...852...96W} for hot DOGs; \citealt{2017MNRAS.464.3431H} for ERQs).

The Ly$\rm \alpha$ emission shows a spatial offset relative to both the UV continuum and H$\rm \alpha$ emission. While the Ly$\rm \alpha$ blue peak shows a slight spatial offset, the Ly$\rm \alpha$ red peak shows a more pronounced spatial offset compared to the blue peak (upper left and right panel of Figure~\ref{fig:Lya_spatial_distribution}). We divide the red peak into three parts along the spatial direction to investigate the anisotropy of the red peak (upper left panel of Figure~\ref{fig:Lya_spatial_distribution}). The red peak seems to be spatially extended and asymmetric, while other emission lines (e.g., \ion{C}{4} and H$\rm \alpha$) do not show any extended emission. The extended Ly$\rm \alpha$ red peak may originate from Ly$\rm \alpha$ photon scattering, with back-scattering occurring on the far side due to disruptions in ISM/CGM structure caused by factors such as outflows, turbulence, or mergers. 

\section{Conclusion and Summary}
In this study, we examined SED properties using photometric data and explored the origin of blue-excess emission using spectroscopic data obtained with Gemini/GMOS and FLAMINGOS-2 for the hyperluminous BlueDOG, ADFS-KMTDOG-102. The key findings from our study of the BlueDOG are summarized as follows:
\begin{itemize}

    \item The estimated stellar mass [log~($M_{*}/M_\odot$)=12.32] and black hole mass [log~$M_{\rm BH}/M_\odot$=10.15], derived from SED fitting and spectrum analysis, confirm that the BlueDOG is an exceptionally massive galaxy. Interestingly, the SED analysis reveals a higher proportion of the old stellar population than the young stellar population, suggesting stellar evolution cumulated from the early universe.

    \item The BlueDOG exhibits a SED shape similar to recently discovered Little Red Dots (LRDs) at \textit{z}$\sim$5--8, aligning with the defining criteria of LRDs. Moreover, the Eddington ratio of the BlueDOG ($\lambda_{\rm Edd}$=0.16) is comparable to that of LRDs ($\lambda_{\rm Edd}$$\sim$0.16), suggesting similar accretion states. However, the SMBH of our sample is 2-3 orders of magnitude more massive.

    \item The rest-frame UV line ratios (e.g., Ly$\rm \alpha$/\ion{N}{5} and \ion{N}{5}/\ion{C}{4}) suggest that these UV lines originate primarily from the central AGN. The BLR appears to be highly metal-enriched. However, the observed nitrogen overabundance may arise from differences in the metal enrichment timescale between nitrogen and carbon rather than reflecting the overall metallicity of the host galaxy.

    \item Synthetic photometry with GMOS spectra shows that the Ly$\rm \alpha$ emission contributes about 40\% of B-band flux, comparable to the continuum contribution. Moreover, the combined contribution of Ly$\rm \alpha$ and \ion{N}{5} accounts for about 60\% of the total flux, indicating that UV emission lines from the AGN dominate the B-band flux. However, the broadband flux boosting due to emission lines does not seem significant on other broadbands since \ion{C}{4} contributes to broadband R-band flux about $\sim$1\% in the UV spectrum of the BlueDOG.

    \item The SED fitting suggests that a recent starburst event, with the sSFR within the past 10\,Myrs, is approximately 6 times higher than that over the past 100\,Myrs. This implies that the blue-excess UV continuum originates from stellar emission rather than the leaked AGN continuum. Similar Balmer break features observed in LRDs further reinforce the starburst scenario. Scattered light modeling also suggests that the observed UV continuum excess in our BlueDOG can be reproduced with a scattered fraction as low as 0.05\% of the intrinsic AGN luminosity. However, since the $L_{3000}$ adopted in the scattered light modeling exceeds the SED-integrated $L_{\mathrm{bol}}$, a scenario in which the UV continuum is entirely produced by scattered light is unlikely to fully explain the observations. Therefore, we cannot conclusively resolve whether the blue excess originates from the star formation or the AGN-scattered light.
    Future deep spectroscopic and polarimetric follow-up observations for our BlueDOG with JWST and other facilities could provide compelling evidence supporting the recent starburst activity or the scattered light from the AGN as the origin of the blue UV continuum, such as identifying spectral features of WR stars (e.g., narrow \ion{He}{2}${\;\lambda\lambda\rm{4686}}$), a Balmer break or high degree of polarization.
    
    \item An AGN model incorporating polar dust provides a more plausible explanation for the BlueDOG's properties, as models without polar dust yield inconsistent results, such as low $E(B-V)$$\sim$0.3 and extremely high SFR of thousands M$_{\odot}$~yr$^{-1}$, which contradict the observed bright FIR fluxes. Moreover, the SMBH mass estimates, derived from extinction-corrected line luminosities using three methods show discrepancies even under a mild polar dust model. 
\end{itemize}

\begin{acknowledgments}
M.K. was supported by the National Research Foundation of Korea (NRF) grant funded by the Korean government (MSIT) (Nos. 2022R1A4A3031306 and RS-2024-00347548). T.N. acknowledges the support by JSPS Kakenhi 23H05441 and 23K17695.

The spectroscopic data in this work were obtained from the K-GMT Science Programs (PID: GS-2017B-Q-18 and GS-2020B-Q-221), supported by the Korean GMT Project and operated by the Korea Astronomy and Space Science Institute.
\end{acknowledgments}

\facilities{Gemini--South (GMOS, FLAMINGOS-2), KMTNet, CTIO (DECam), VISTA, WISE, Spitzer (IRAC, MIPS), AKARI (FIS) and Herschel (PACS, SPIRE)}
\software{Astropy \citep{astropy:2013, astropy:2018, astropy:2022}, CIGALE-2022.1 \citep{2019A&A...622A.103B, 2022ApJ...927..192Y}, NumPy \citep{harris2020array}, SciPy \citep{2020SciPy-NMeth}, Pyphot \citep{zenodopyphot}, Matplotlib \citep{Hunter:2007}}

\bibliography{bluedog}{}
\bibliographystyle{aasjournal}



\end{document}